\documentclass[12pt,preprint, showpacs]{revtex4}

\usepackage[brazil, english]{babel}
\usepackage[utf8]{inputenc}
\usepackage{graphicx}
\usepackage{epsfig} 
\usepackage{amssymb}
\usepackage{amsmath}
\usepackage{slashed}

\graphicspath{%
    {converted_graphics/}
    {/}
}
\begin{document}

\title{Pomeron and Odderon Regge Trajectories from a Dynamical Holographic Model}
\author{Eduardo Folco Capossoli$^{1,2,}$}
\email[Eletronic address:]{educapossoli@if.ufrj.br}
\author{Danning Li$^{3,}$}
\email[Eletronic address:]{lidn@itp.ac.cn}
\author{Henrique Boschi-Filho$^{1,}$}
\email[Eletronic address: ]{boschi@if.ufrj.br}  
\affiliation{$^1$Instituto de F\'{\i}sica, Universidade Federal do Rio de Janeiro, 21.941-972 - Rio de Janeiro-RJ - Brazil \\
 $^2$Departamento de F\'{\i}sica, Col\'egio Pedro II, 20.921-903 - Rio de Janeiro-RJ - Brazil \\ 
 $^3$ Institute of Theoretical Physics, Chinese Academy of Science (ITP, CAS), 100190 - Beijing - China}

\begin{abstract}
In this work we use  gauge/string dualities and a dynamical model that takes into account dynamical corrections to the metric of the anti de Sitter space due to a quadratic dilaton field and calculate the masses of even and odd spin glueball states with $P=C=+1$, and  $P=C=-1$, respectively. Then we construct the corresponding Regge trajectories which are associated with the pomeron for even states with $P=C=+1$, and with the odderon for odd states with $P=C=-1$. We compare our results with those coming from experimental data as well as other models. 
\end{abstract}


\maketitle

\section{Introduction}

Quantum Chromodynamics (QCD) is a non-abelian gauge theory that describes the strong interactions between quarks and gluons. Despite the fact that gluons do not carry electric charges, they have color charge. Due to this fact, they can couple to each other and form bound states  called glueballs which have not been detected so far, becoming itself a great QCD quest. 
At high energies the QCD coupling is small and perturbative methods work well. At low energies where bound states form (hadronization) QCD coupling is large and perturbative methods fail. 

Various current studies deal with glueball issues from both experimental and theoretical points of view \cite{Qiao:2014vva, Chuinard:2015sva}. On the theoretical side one can see several approaches such as lattice QCD, the flux tube model, MIT bag model, Coulomb Gauge model and QCD Sum Rules \cite{Qiao:2014vva}. There is also a novel approach based on holography or AdS/CFT or Anti de Sitter/Conformal Field Theory correspondence \cite{Maldacena:1997re, Gubser:1998bc, Witten:1998qj, Witten:1998zw, Aharony:1999ti} to circumvent the difficulty of  non-perturbative QCD at low energies. 

Motivated by last ten years of efforts based on AdS/CFT correspondence to investigate glueball states \cite{Polchinski:2001tt, Polchinski:2002jw, BoschiFilho:2002vd, BoschiFilho:2002ta, deTeramond:2005su, Brunner:2015yha, Bellantuono:2015fia, Brower:2014wha, Li:2013oda, Capossoli:2015ywa, Capossoli:2013kb, Colangelo:2007pt, BoschiFilho:2005yh, Chen:2015zhh, Ballon-Bayona:2015wra, Brower:2006ea, Brower:2008cy}  the main objective of this work is to calculate the masses for both even and odd spin glueballs and obtain the Regge trajectories related to the pomeron and the odderon. We use a dynamical holographic model, taking into account dynamical corrections to the anti de Sitter (AdS) space metric due to a quadratic dilaton field. This is the first calculation of high spin glueballs $(J>2)$ with a dynamical model. This allows us to solve open some questions on the Regge trajectories for the pomeron and the odderon. 

The AdS/CFT correspondence or duality is a powerful tool to tackle non-perturbative Yang-Mill theories. This  duality relates a conformal Yang-Mills theory with the symmetry group $SU(N)$ for very large $N$ and extended supersymmetry $({\cal N} = 4)$  with a $IIB$ superstring theory in a curved space, known as anti de Sitter space, or $AdS_5 \times S^5$. At low energies string theory is represented by an effective supergravity theory, due to this reason the AdS/CFT is also known by gauge/gravity duality.

After breaking the conformal symmetry one can build phenomenological models that describe approximately QCD. These models are known as AdS/QCD models.

In order to deal with conformal symmetry breaking the works \cite{Polchinski:2001tt, Polchinski:2002jw, BoschiFilho:2002ta, BoschiFilho:2002vd} have done some important progress with this issue. In these  works, emerged the idea of the hardwall model, which meant that a hard cutoff was introduced at a certain value $z_{max}$ of the holographic coordinate $z$ and it was considered a slice of $AdS_5$ space in the region $0 \leq z  \leq z_{max}$.

Another holographic model to break the conformal invariance in the boundary theory, and make it an effective theory of QCD, is called  the softwall model (SWM). This model  introduces an exponential factor in the action related to a dilatonic filed that represents a soft IR cutoff. The SWM was proposed in \cite{Karch:2006pv} to study  vector mesons, and subsequently extended to glueballs \cite{Colangelo:2007pt}. One interesting property of the SWM is to provide linear Regge trajectories. In the ref. \cite{Capossoli:2015ywa} it was shown that SWM does not work properly for calculation of scalar glueball and its radial excitation masses or higher spin glueball states. Due to this, one cannot find satisfactory Regge trajectories for glueballs from the SWM consistent with the literature.

In this work, our principal aim is to calculate Regge trajectories for both even and odd higher spin glueballs using a dynamical version of the SWM, i.e, imposing that the dilaton field became dynamical satisfying the Einstein equations in five dimensions \cite{Li:2013oda}. The Regge trajectories obtained for both even and odd spin glueballs, related to the pomeron and the odderon, respectively, are in good agreement with available data. 


\section{The dynamical holographic  Model}

The holographic dynamical softwall (DSW) model that we are going to consider has a metric structure which is consistently solved from Einstein's equation. To obtain the metric solution we write a $5D$ action for the graviton-dilaton coupling in the string frame:
\begin{equation}\label{acao_corda}
S = \frac{G_5^{-1}}{16 \pi } \int d^5 x \sqrt{-g_s} \; e^{-2\Phi(z)} ({\cal R}_s + 4 \partial_M \Phi \partial^M \Phi - V^s_G(\Phi))
\end{equation}
\noindent where $G_5$ is the Newton's constant in five dimensions,  the dilaton field $\Phi$ is given by $\Phi = kz^2$, where $k \sim \Lambda^2_{QCD}$ and $V_G$ is the dilatonic potential. 
The metric tensor in 5-dimensional space has the following form $g^s_{mn} = b^2_s(z)(dz^2 + \eta_{\mu \nu}dx^\mu dx^\nu)$ with  $0 \leq z \leq \infty$, $m,n = 0,1,2,3,4\; , \;\mu, \nu = 0,1,2,3 \;{\rm and} \; \eta_{\mu \nu} =$ diag $(-1, 1, 1, 1)$ and $b_s(z) \equiv e^{A_s(z)}$. All of these parameters are in the string frame. 
The precise form of the metric will be defined solving the equations of motion and finding explicitly the expression for $A_s(z)$. Actually, it is easier to solve this problem in the Einstein frame. 

Performing a Weyl rescaling,  from the string frame to the Einstein frame, such that
$b_E (z) = b_s(z)e^{-\frac{2}{3}\Phi(z)} = e^{A_E(z)}$ and $A_E(z) = A_s(z) - \frac{2}{3}\Phi(z)$, 
 one can rewrite the action (\ref{acao_corda}) as:
\begin{equation}\label{acao_einstein}
S = \frac{G_5^{-1}}{16 \pi } \int d^5 x \sqrt{-g_E} \; ({\cal R}_E -\frac{4}{3} \partial_m \Phi \partial^m \Phi - V^E_G(\Phi))
\end{equation}
\noindent where $g^E_{mn} = g^s_{mn}e^{-\frac{2}{3}\Phi}\; {\rm and} \; V^E_G = e^{\frac{4}{3}\Phi}V^s_G$. 
Varying this action, one can obtain the equations of motion: 
\begin{eqnarray}\label{eq_mov_e_2_1}
 & -A''_E + A'^2_E - \frac{4}{9}\Phi'^2  = 0 \;;\\ 
 & \Phi'' + 3A'_E \Phi' - \frac{3}{8}e^{2A_E}\partial_\Phi V^E_G(\Phi) = 0 \,.
 \label{eq_mov_e_2_2}
\end{eqnarray}

\noindent Solving these equations for the quadratic dilaton $\Phi = kz^2$, one finds: 
\begin{eqnarray}\label{sol_eq_mov_e_2_1}
 A_E(z) &=& \log{\left( \frac{R}{z} \right)} - \log{(_0F_1(5/4, \frac{\Phi^2}{9}))}\;;
\\
 V^E_G(\Phi) &=& -\frac{12}{R^2} ~ _0F_1(1/4, \Phi^2/9)^2 \nonumber \\
& &+ \frac{16}{3 R^2} ~ _0F_1(5/4, \Phi^2/9)^2 \Phi^2\,,\label{sol_eq_mov_e_2_2}
\end{eqnarray}
\noindent where $R$ is the AdS radius. 

Going back to string frame, one can write the $5D$ action for the scalar glueball as  \cite{Colangelo:2007pt}:
\begin{equation}\label{acao_ori_soft}
S = \int d^5 x \sqrt{-g_s} \; \frac{1}{2} e^{-\Phi(z)} [\partial_M {\cal G}\partial^M {\cal G} + M^2_{5} {\cal G}^2]
\end{equation}
\noindent and the equations of motion are:
\begin{equation}\label{eom_1}
\partial_M[\sqrt{-g_s} \;  e^{-\Phi(z)} g^{MN} \partial_N {\cal G}] - \sqrt{-g_s} e^{-\Phi(z)} M^2_{5} {\cal G} = 0\,.
\end{equation}
Representing the scalar field through a $4d$ Fourier transform ${\cal \tilde{G}}(q,z)$ and performing a change of function ${\cal \tilde{G}} = \psi (z) e^{\frac{B(z)}{2}}$, where $B(z) = \Phi(z) - 3A_s(z) $, one gets the following $1d$ Schr\"odinger-like equation
\begin{eqnarray}\label{equ_7_new1}
 - \psi''(z) + V(z) \psi(z) = (- q^2 )\psi(z)\,,
\end{eqnarray}
\noindent where the effective potential is given by 
\begin{eqnarray}\label{V(z)}
 V(z) =  k^2 z^2 + \frac{15}{4z^2}  - 2k 
 +\left( \frac{M_{5}R}{z}\right)^2  e^{4kz^2/3} {\cal A}^{-2}  \,,
\end{eqnarray}
with ${\cal A}$ = $_0F_1(5/4, \frac{\Phi^2}{9})$. The normalizable solutions of Eq. \eqref{equ_7_new1} correspond to a discrete spectrum of $4d$ masses with the identification $q^2 = -m_n^2$. 
This equation was solved numerically in \cite{Li:2013oda}  for the scalar glueball $0^{++}$ and its radial (spin 0)  excitations and the masses found are compatible with those from lattice QCD.

From the AdS/CFT  dictionary one knows how to relate the operator in the super Yang-Mills theory with fields in the $AdS_{5} \times S^5$ space. 
In the case of dual higher spin fields we consider a symmetric traceless spin $J$ field and the relation between the conformal dimension $\Delta$ and the AdS mass is 
\begin{equation}\label{hsp}
M^2_{5}R^2 =  \Delta (\Delta - 4) - J \,; \qquad (J=0, 1, 2, 3, \cdots)
\end{equation}
and the effective potential reads
\begin{equation}\label{ve}
V_J(z) = k^2 z^2 + \frac{15}{4z^2}  - 2k +  \frac{\Delta (\Delta - 4) - J}{z^2} \, e^{4kz^2/3} {\cal A}^{-2}\,. 
\end{equation}


\section{Even spins and the pomeron.} 

In the perturbative approach, the pomeron is identified with the leading twist 2 trajectory $\Delta=  J +2$ (see, e. g., \cite{Brower:2006ea} and refs. therein). Note that for twist two operator the relation \eqref{hsp} gives $M^2=J^2-J-4$ as shown in \cite{Karch:2006pv}. 
So, let us start this study with this assumption. Using the DSW model discussed in the previous section, one obtains complex masses for the glueball states $0^{++}$ and $2^{++}$ which does not allow us to find a Regge trajectory for the pomeron. However, in the non-perturbative regime higher twist operators also contribute to the scattering amplitudes related to the pomeron  \cite{Donnachie:2002en}. 
Furthermore, in ref. \cite{Meyer:2004jc} it was argued that the glueball state 0$^{++}$ does not belong to the pomeron trajectory, but to another one with lower intercept. This does not agree with \cite{Donnachie:2002en} where the 0$^{++}$ state is taken into account to the pomeron trajectory. In order to investigate these problems, we are going to consider trajectories including and excluding the 0$^{++}$ state  and with higher twist.

So now we consider twist four operators for a pure super Yang-Mills theory defined on the 4D boundary, such that $\Delta=  J +4$ and compute the masses for even glueball states using eqs. \eqref{hsp} and \eqref{ve}. 
The scalar glueball state $0^{++}$ is represented by the operator ${\cal O}_4$, and it can be written as ${\cal O}_4 = Tr(F^2) = Tr(F^{\mu\nu}F_{\mu \nu})$.
For higher spin glueballs we insert symmetrized covariant derivatives in a given operator with spin in order to raise the total angular momentum.
Then, one obtains ${\cal O}_{4 + J} = FD_{\lbrace\mu1 \cdots} D_{\mu J \rbrace}F$, with conformal dimension $\Delta = 4 + J$ and spin $J$.

To calculate the masses for the higher spin glueball states and get the Regge trajectory related to the pomeron, one has to solve the eq. (\ref{equ_7_new1}) numerically with the effective potential \eqref{ve}. 
The masses found  are shown in Table \ref{t1}. 

\begin{table}[!h]

\centering
\begin{tabular}{|c|c|c|c|c|c|c|c|}
\hline
 &  \multicolumn{6}{c|}{Glueball States $J^{PC}$}  & \\  
\cline{2-7}
 & $0^{++}$ & $2^{++} $ & $4^{++}$ & $6^{++}$ & $8^{++}$ & $10^{++}$  & $ k $ \\
\hline \hline
 $\; m_n \; $ &\, 0.51\, &\, 2.03 \,&\, 3.23 \,& \, 4.40 \, &\, 5.56 \,&\, 6.71 \, & \,  0.10 \,   \\ \hline 
\end{tabular} 
\caption{Masses $\; m_n \; $    expressed in GeV for the glueball states $J^{PC}$ with even $J$ as the eigenstates of Eq. (\ref{equ_7_new1}) with the potential (\ref{ve}) for  $k= 0.10$ GeV$^2$.}
\label{t1}
\end{table}

Regge trajectories are an approximate linear relation between total angular momenta $(J)$ and the square of the masses $(m)$, such that $J(m^2) \approx \alpha_0 + \alpha' m^2$ with $\alpha_0$ and $\alpha'$ constants. 
One can obtain Regge trajectories for the pomeron using data from Table \ref{t1} once they are associated with even spin glueballs. 

\begin{figure}[h] 
  \centering
  \includegraphics[scale = 0.18,angle=-90]{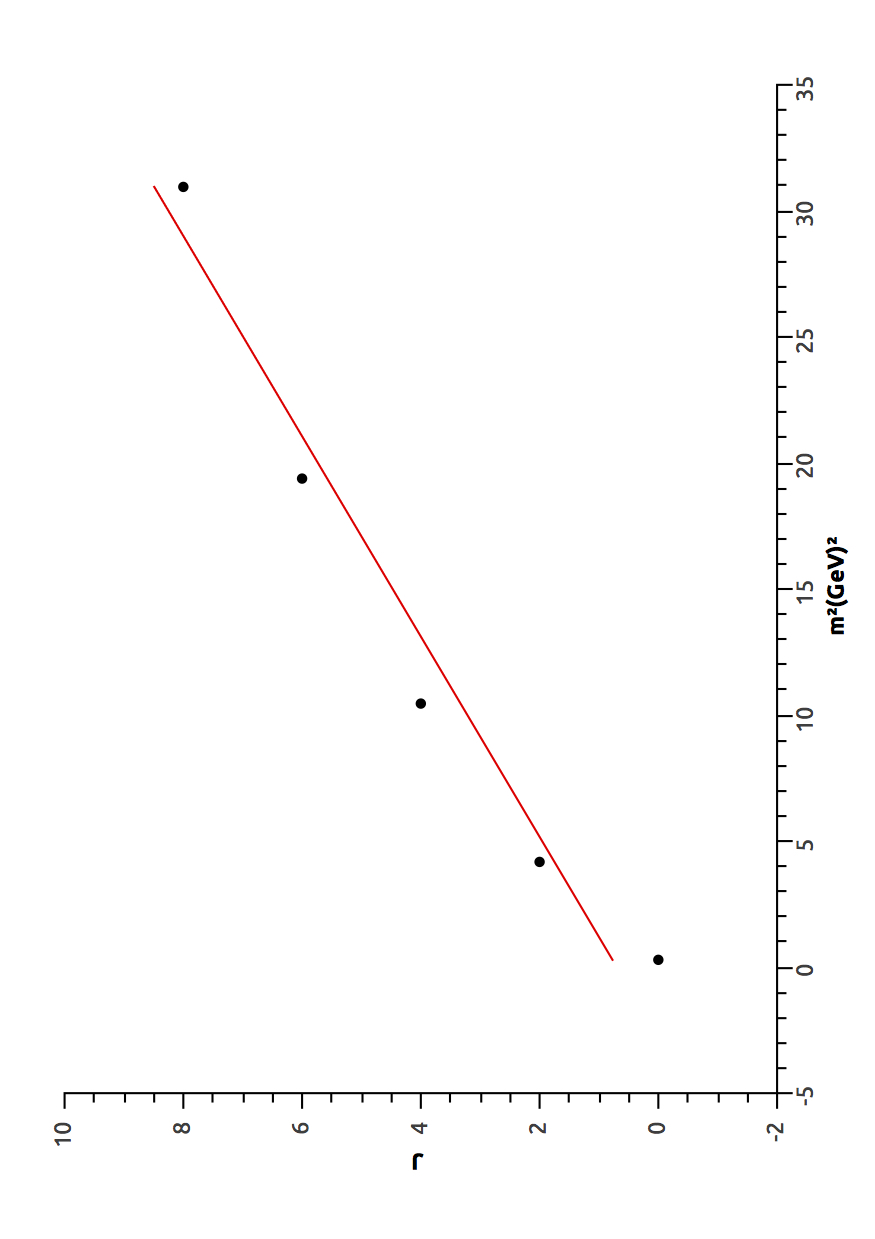}\includegraphics[scale = 0.18,angle=-90]{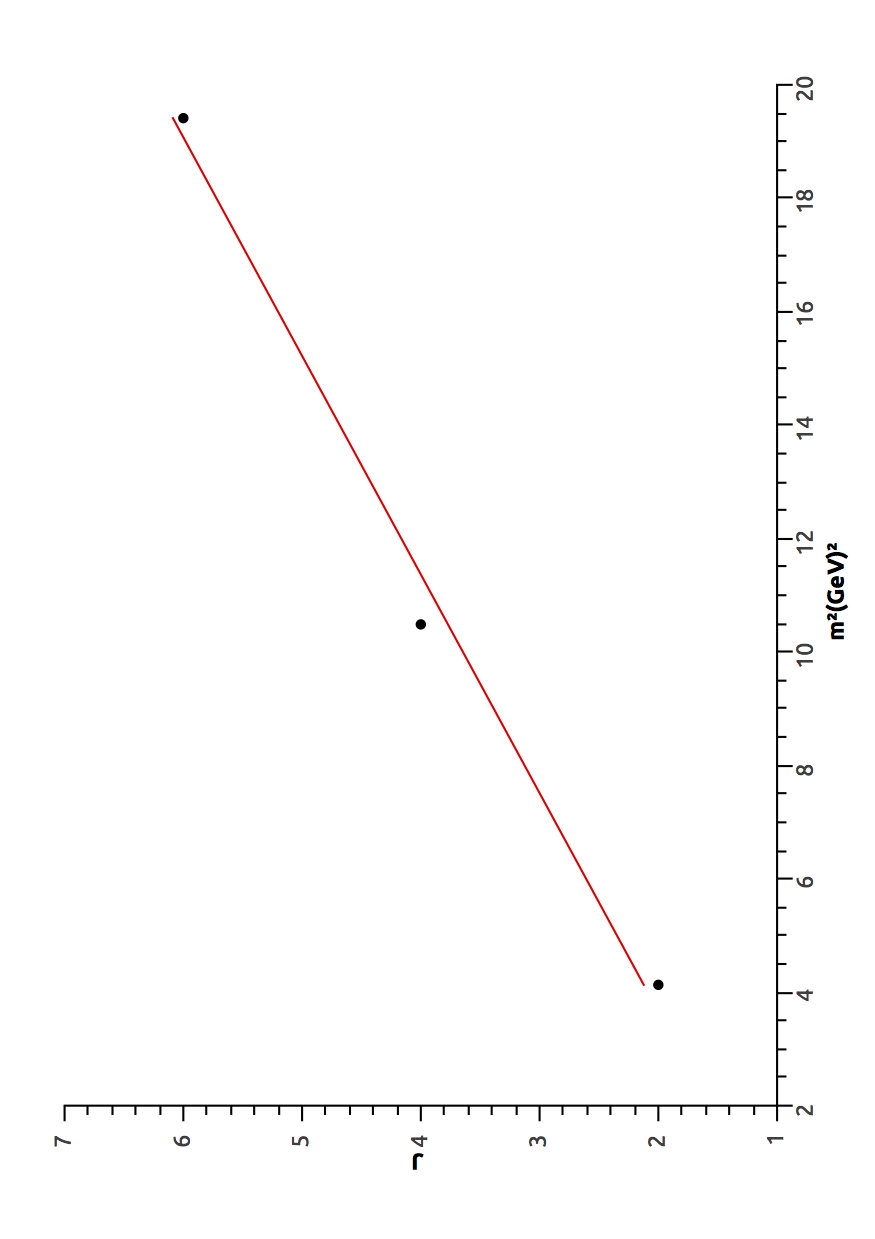}
\caption{Regge trajectories for the pomeron, from data of Table \ref{t1} with $k=0.1$ GeV$^2$. Left panel: glueball states $0^{++}, 2^{++}, 4^{++}, 6^{++},$ and $8^{++}$ and trajectory given by eq. \eqref{rpsd}. Right panel: glueball states $2^{++}, 4^{++}$ and  $6^{++}$ and trajectory given by eq. \eqref{rpszsd}.}
\label{DSWpomeron}
\end{figure}

For instance, a Regge trajectory can obtained from eq.(\ref{equ_7_new1}) for the pomeron using table \ref{t1} and $k= 0.10$ GeV$^2$, excluding the state $10^{++}$, and is presented bellow:
\begin{equation}\label{rpsd}
J(m^2) \approx (0.72 \pm 0.49) + (0.25 \pm 0.02) m^2
\end{equation}
The errors come from the linear fit. This Reege trajectory for the pomeron is represented in Figure \ref{DSWpomeron} (left panel) and is in agreement with the one presented in \cite{Landshoff:2001pp}.

Choosing another set of states, for exemple, $2^{++}, 4^{++}, 6^{++}$, from Table \ref{t1} also with $k$ = 0.10 GeV$^2$, one finds the following Regge trajectory:
\begin{equation}\label{rpszsd}
J(m^2) = (1.06 \pm 0.33) + (0.26 \pm 0.02) m^2
\end{equation}
\noindent which is represented in Figure \ref{DSWpomeron} (right panel) and is in excellent agreement with \cite{Landshoff:2001pp} and also with \cite{Meyer:2004jc} where the  state $0^{++}$ was excluded.

\begin{figure}[h] 
  \centering
  \includegraphics[scale = 0.55]{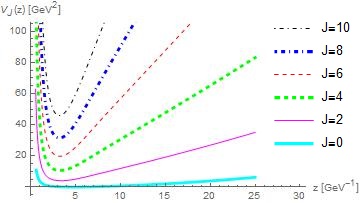}   \includegraphics[scale = 0.55]{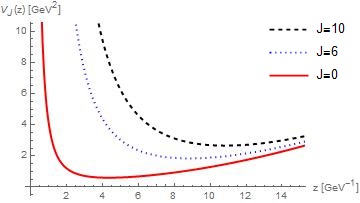} 
\caption{Plots of the effective potentials $V_J(z)$ against the holographic coordinate $z$ for some values of even spins. Left panel: $J=0, 2, \cdots, 10$ and effective potential given by Eq. (\ref{ve}), in the DSW model. Right panel: $J=0$, $J=6$, and $J=10$  in the SWM.}
\label{plot1}
\end{figure}

We plot in Figure \ref{plot1} (left panel) the shape of the effective potential $V_J(z)$ in the DSW model for different values of the spin $J$. 
One can see that in the UV limit $(z \rightarrow 0)$ all plots have a similar behavior. But in the IR region $(z \rightarrow \infty)$ they differ clearly for different spins. The higher the spin of the glueball, the higher is the slope of the effective potential at large $z$. Also the minimum of the potential increases with the spin of the glueball. 

For comparison, we show in Figure \ref{plot1} (right panel) the corresponding effective potentials for the SWM extended for high spins \cite{Capossoli:2015ywa}. These effective potentials also increase fast for small $z$, but they increase slowly for large $z$, in contrast with the holographic dynamical model shown in Figure \ref{plot1} (left panel). 

\section{Odd spins and the odderon.} 

For odd spin glueballs, the operator ${\cal O}_6$ that describes the glueball state $1^{--}$ is given by ${\cal O}_{6} =SymTr\left( {\tilde{F}_{\mu \nu}}F^2\right)$
 and after the insertion of symmetrized covariant derivatives one obtains ${\cal O}_{6 + J} = SymTr\left( {\tilde{F}_{\mu \nu}}F D_{\lbrace\mu1 \cdots} D_{\mu J \rbrace}F\right)$, with conformal dimension $\Delta = 6 + J$ and spin $1+J$.

To calculate the masses for the higher spin glueball states and get the Regge trajectory related to the odderon, we  solve the eq. (\ref{equ_7_new1}) with the effective potential \eqref{ve} numerically  for  odd $J$. 
The masses eigeinstates found for odd glueball states are shown in Table \ref{t2}. 

\begin{table}[h]

\centering
\begin{tabular}{|c|c|c|c|c|c|c|c|}
\hline
 &  \multicolumn{6}{c|}{Glueball States $J^{PC}$}  & \\  
\cline{2-7}
 & $1^{--}$ & $3^{--} $ & $5^{--}$ & $7^{--}$ & $9^{--}$ & $11^{--}$  & $ k $ \\
\hline \hline
$\; m_n \; $                    
&\, 2.77 \, &\, 3.91 \,&\, 5.05 \,& \, 6.19 \, &\, 7.33 \,&\, 8.47 \, & \, 0.10 \,  \\ \hline
\end{tabular}
\caption{ Masses $\; m_n \; $    expressed in GeV for the glueball states $J^{PC}$ with odd $J$  solving Eq. (\ref{equ_7_new1}) with the potential (\ref{ve}) for $k= 0.10$ GeV$^2$.} 
\label{t2}
\end{table}

Regge trajectories for the odderon can be obtained using data from Table \ref{t2} since they are associated with odd spin glueballs. 

\begin{figure}[h] 
  \centering
  \includegraphics[scale = 0.18,angle=-90]{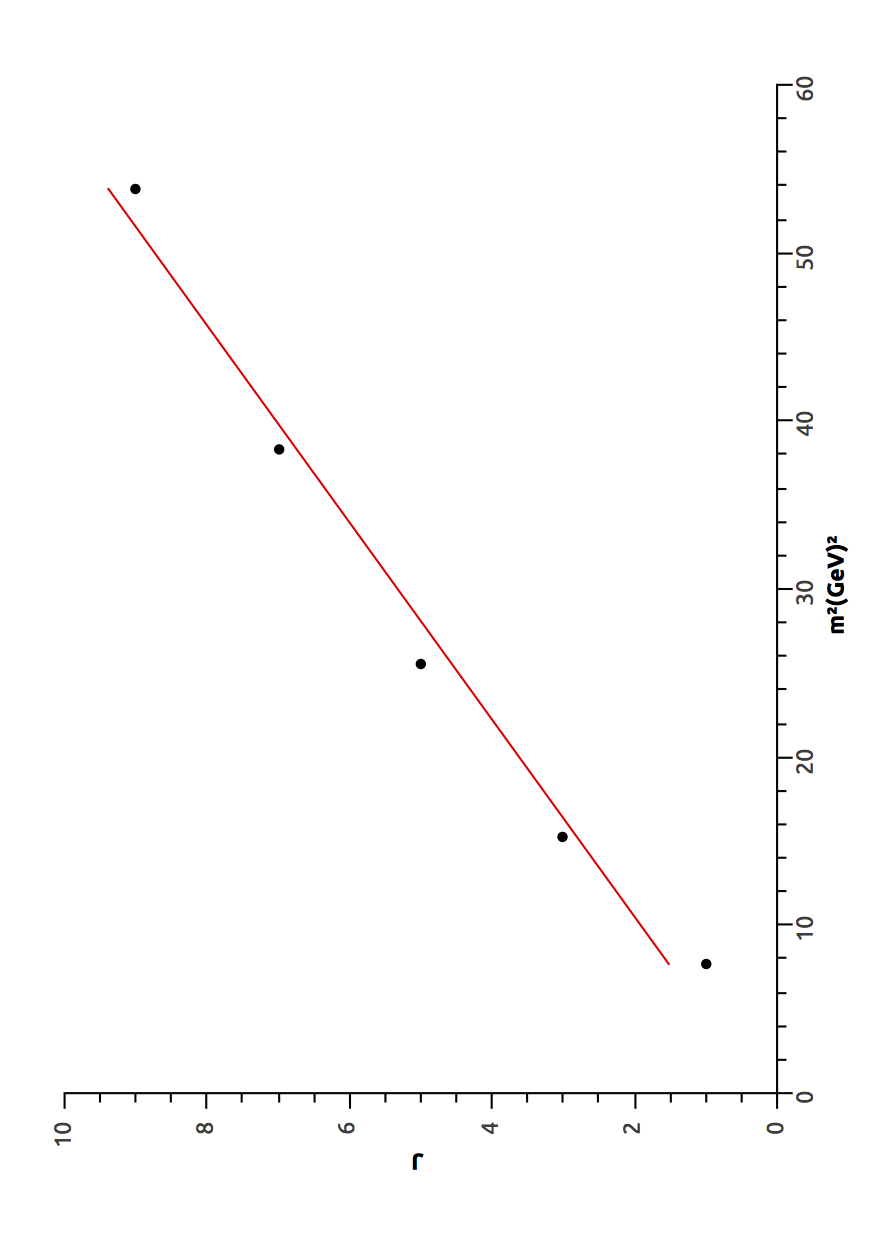} \includegraphics[scale = 0.18,angle=-90]{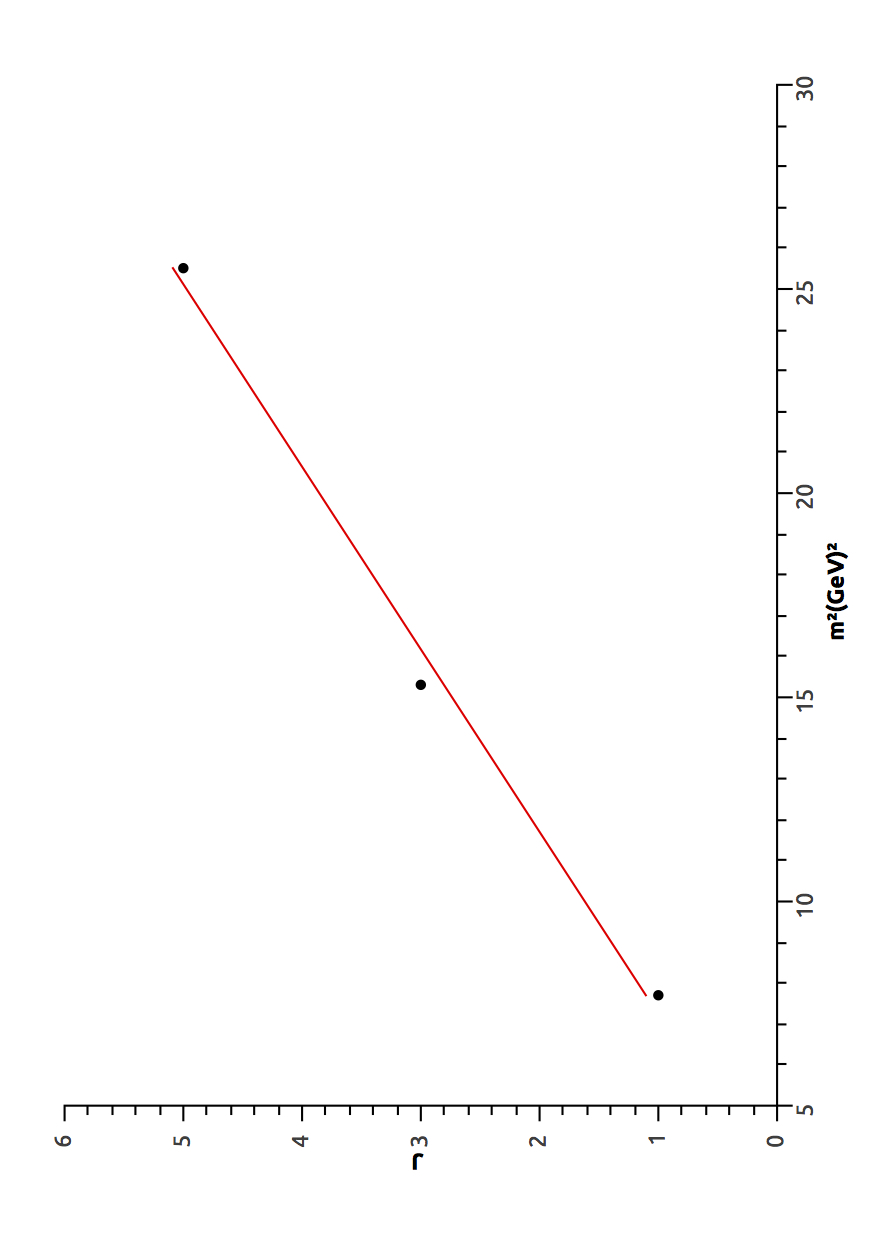}
\caption{Regge trajectories for the odderon, from data of Table \ref{t2} with $k=0.1$ GeV$^2$. Left panel: glueball states $ 1^{--}, 3^{--}, 5^{--}, 7^{--} $,  $9^{--}$ and trajectory given by eq. \eqref{rpso}. Right panel: glueball states $ 1^{--}, 3^{--}, 5^{--}$ and trajectory given by eq. \eqref{rpso2}.}
\label{DSWodderon}
\end{figure}

For the odderon, using table \ref{t2} and $k= 0.10$ GeV$^2$, excluding the state $11^{--}$, the Regge trajectory is presented bellow:
\begin{equation}\label{rpso}
J(m^2) \approx (0.20 \pm 0.43) + (0.17 \pm 0.01) m^2
\end{equation}
The errors come from the linear fit. This Reege trajectory for the odderon is represented in Figure \ref{DSWodderon} (left panel) and is in agreement with  \cite{LlanesEstrada:2005jf} within the non-relativistic constituent model.

Choosing another set of states, for exemple, $1^{--}, 3^{--}, 5^{--}$, from Table \ref{t2} also with $k$ = 0.10 GeV$^2$, one finds the following Regge trajectory:
\begin{equation}\label{rpso2}
J(m^2) = (-0.60 \pm 0.33) + (0.22 \pm 0.01) m^2
\end{equation}
which is compatible  with  \cite{LlanesEstrada:2005jf} within the relativistic many-body model. This Reege trajectory for the odderon is represented in Figure \ref{DSWodderon} (right panel) and the errors also come from the linear fit.

Another set with  $3^{--}, 5^{--}, 7^{--}$, from Table \ref{t2} also with $k$ = 0.10 GeV$^2$ gives 
\begin{equation}\label{rpso3}
J(m^2) = (0.44 \pm 0.32) + (0.17 \pm 0.01) m^2
\end{equation}
which is compatible with the non-relativistic constituent model \cite{LlanesEstrada:2005jf} and excludes the state  
$1^{--}$.

\begin{figure}[h] 
  \centering
  \includegraphics[scale = 0.55]{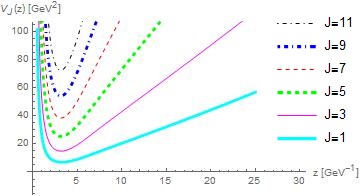}   \includegraphics[scale = 0.55]{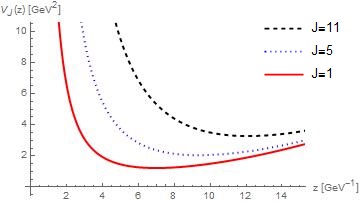} 
\caption{Plots of the effective potentials $V_J(z)$ against the holographic coordinate $z$ for some values of odd spins. Left panel:  $J=1, 3, \cdots, 11$ and effective potential given by Eq. (\ref{ve}), in the DSW model. Right panel:  $J=1$, $J=5$, and $J=11$ in the SWM.}
\label{plot3}
\end{figure}

Figure \ref{plot3} (left panel) represents the effective potentials in the DSW model for various odd spin glueball states. 
For comparison, we show in Figure \ref{plot3} (right panel) the corresponding effective potentials for the usual SWM extended for high spins \cite{Capossoli:2015ywa}. These effective potentials also increase fast for small $z$, but they increase slowly for large $z$, in contrast with the holographic dynamical model shown in Figure \ref{plot3} (left panel). 

\section{conclusions}

We used a dynamical holographic softwall model to obtain even and odd spin glueball mass spectra and achieve the related Regge trajectories associated with the pomeron and the odderon, respectively. These trajectories are in good  agreement with those found in \cite{Capossoli:2015ywa, 
Capossoli:2013kb, LlanesEstrada:2005jf, BoschiFilho:2005yh, Landshoff:2001pp, Meyer:2004jc, Ballon-Bayona:2015wra}.  Besides, this is the first obtention of these Regge trajectories through a dynamical model. 

In particular, it has been shown in \cite{Capossoli:2015ywa} that the SWM could not lead to the expected Regge trajectories for the pomeron or the odderon. This problem is overcome in this work by the use of the DSW model. 

The masses obtained in this work can be compared with the ones coming from some phenomenological models. First, the masses found here from the DSW model are higher than those found from the SWM \cite{Capossoli:2015ywa}, if one chooses the same value of the free parameter $k$. This fact can be understood comparing the plots $V_J(z)\; \times\;  z$ for these two models. Figures \ref{plot1} and \ref{plot3} (both left panels) represent  the effective potential for the DSW model for some even and odd spin glueball states while  Figures \ref{plot1}  and \ref{plot3} (both right panels) represent the corresponding potentials for the softwall model. Since the minima of the effective potentials for the DSW model are higher than those of the SWM, the corresponding masses eigenstates for the DSW model are higher then those from the SWM. 

Second, the masses found from the DSW model are similar to those found in refs. \cite{Capossoli:2013kb, BoschiFilho:2005yh} within the holographic hardwall model \footnote{This identification works better for the hardwall model with Neumann boundary condition, which has good agreement with experimental data.}. This fact can be understood also comparing Figures \ref{plot1} and \ref{plot3} (left panels) for the DSW model with  Figures \ref{plot1}  and \ref{plot3} (right panels) for the SWM. One can see that in Figures \ref{plot1} and \ref{plot3} (let panels) the effective potentials increase with larger slopes for each spin than those in Figures \ref{plot1}  and \ref{plot3} (right panels). This means that the dynamical corrections of the SWM produce barriers for the effective potential similar to those of the phenomenological hardwall model. 

In ref. \cite{Meyer:2004jc}  it was argued that the state $0^{++}$ does not belong to the pomeron's Regge trajectory. Our results on the Regge trajectories for the pomeron, showed in Eqs. (\ref{rpsd}) and  (\ref{rpszsd}) are not conclusive in this regard. In Eq. (\ref{rpsd}) we obtain a Regge trajectory for the pomeron including the state $0^{++}$ which is agreement with \cite{Landshoff:2001pp} and \cite{Ballon-Bayona:2015wra}. On the other side, in Eq. (\ref{rpszsd}) we obtain a Regge trajectory for the pomeron excluding the state $0^{++}$ which is also compatible with experimental data \cite{Landshoff:2001pp}. Note, however, that this is our best fit for the pomeron trajectory.

In the case of the odderon, different Regge trajectories were found in \cite{LlanesEstrada:2005jf} corresponding  to a relativistic and a non-relativistic models. We found compatible results within the DSW model for both the relativistic and non-relativistic predictions for the odderon as presented in \cite{LlanesEstrada:2005jf}. 
Also, in ref. \cite{LlanesEstrada:2005jf} it was argued that the state $1^{--}$ should not belong to the odderon Regge trajectory. This conclusion is not supported by our results with the DSW model for the exclusion of the $1^{--}$ state, since we found good trajectories for the odderon with and without the glueball state $1^{--}$.

\begin{acknowledgments}
H.B.-F. is partially supported by CNPq and E.F.C. by CNPq and FAPERJ, Brazilian agencies. D.L. is supported by the China Postdoctoral Science Foundation. 
\end{acknowledgments}

\end{document}